\documentclass[nofootinbib,notitlepage,superscriptaddress,10pt,aps,prd,twocolumn]{revtex4-1}

\usepackage{amsmath,mathtools}
\usepackage{tensor}
\usepackage{float}
\usepackage{amssymb}
\usepackage{graphicx}
\usepackage{verbatim}
\newcommand{\leri}[1]{\left(#1\right)}

\newcommand{\abs}[1]{\left|#1\right|}
\begin{document}

\title{Implications of the Holst term in a $f(R)$ theory with torsion}
\author{Flavio Bombacigno}
\email{flavio.bombacigno@uniroma1.it}
\affiliation{Physics Department, ``Sapienza'' University of Rome, P.le Aldo Moro 5, 00185 (Roma), Italy}
\author{Giovanni Montani}
\email{giovanni.montani@enea.it}
\affiliation{ENEA, FSN-FUSPHY-TSM, R.C. Frascati, Via E. Fermi 45, 00044 Frascati, Italy.\\
Physics Department, ``Sapienza'' University of Rome, P.le Aldo Moro 5, 00185 (Roma), Italy}

\begin{abstract}
We analyze a modified $f(R)$ theory of gravity 
in the Palatini formulation, when an Holst term endowed with a dynamical 
Immirzi field is 
included. We study the basic features of the model, 
especially in view of eliminating the torsion field via the 
Immirzi field and the scalar-tensor degrees of freedom of the 
$f(R)$ model. 
The main task of this study is the investigation of the morphology of the gravitational wave polarization when their coupling to a circle of test particles is considered. 
We first observe that the dynamics 
of the scalar mode of the $f(R)$ Lagrangian is frozen 
out, since its first order term identically vanishes. 
This allows a detailed characterization of the linearized theory, which outlines the emergence of a modified Newtonian potential in the static limit, and when time independence is relaxed a standard gravitational wave plus the scalar wave associated to the Immirzi field. 
Investigating the effect of the coupling of this 
scalar-tensor wave on a circle of test particles, we 
arrive to define two effective gravitational polarizations, 
corresponding to an equivalent phenomenological wave, 
whose morphology is anomalous with respect the standard case 
of General Relativity. 
In fact, the particle circle suffers modifications as 
it was subjected to modified plus and cross modes, 
whose specific features depend on the model free parameters and are, in principle, detectable via a data analysis procedure. 
\end{abstract}

\maketitle

\section{Introduction}
The idea that the dynamics of the gravitational field is not uniquely fixed by the implementation of the Einstein equations 
is a well established theme in literature \cite{Bergmann:1968ve,Schmidt:2006jt,Sotiriou:2008rp,Nojiri:2010wj} and, in the last three decades, it has acquired a clear physical motivation in the necessity to account for 
exotic phenomena, like dark energy and dark matter \cite{Starobinsky,Whitt:1984pd,Peebles:2002gy}. \\In fact, in addition to the request that 
the modified theories of gravity be able to remove the existence of curvature singularity, overall the Big-Bang singularity, now they are derived with the non-trivial aim of explaining for new physics \cite{Cognola:2007zu,Lecian:2008kg,Elizalde:2010ts,Nojiri:2017ncd,Odintsov:2017qif,Capozziello:2018wul}. 
\\Actually, dark energy seems a purely dynamical effect in the late Universe expansion and it stands as one of the most 
promising candidate for being addressed via 
modified gravity effects. 
Among the infinite variety of possible restatements of Einsteinian dynamics the so-called $f(R)$ model \cite{,Nojiri:2017ncd} stands for its simultaneous generality and simplicity: it is well-known its equivalence with scalar-tensor theory \cite{ST}, whose dynamics is easily accounted. Nonetheless, the $f(R)$ gravity outlines a peculiar feature in a somewhat break down of the 
equivalence between the metric and Palatini formulation \cite{Pal}. This interesting and, to some extent, puzzling feature is suggesting that a coherent Palatini formulation of the theory requires the introduction of a torsion field \cite{Hehl:1976kj,Shapiro:2001rz,Cartan1,Cartan2}
\textit{ab initio} in the Lagrangian model. 
This idea was successifully pursued in \cite{Bombacigno:2018tbo}, where a Nieh-Yan term \cite{NY1,NY2,Mercuri:2007ki} is included in the modified Lagrangian, 
also in the presence of an Immirzi field \cite{Calcagni:2009xz,TorresGomez:2008fj,Cianfrani:2009sz,Bombacigno:2016siz}
\\ The implications for the morphology of the gravitational waves and possible phenomenological signature of 
the theory have been also discussed there. 
\\ Furthermore, the cosmological implementation of the proposed theory of modified gravity on a cosmological setting has been also addressed, for viable form of the $f(R)$, in \cite{Bombacigno:2018tyw}. 
In particular, in the case of an exponential Lagrangian in the non-Riemannian Ricci scalar, a feature of dark energy emerges in the late Universe expansion. Despite the Nieh-Yan term is an elegant choice 
(we recall that for a constant Immirzi parameter, it is a topological term) and even if it provides a simplification allowing to reduce the original Lagrangian by eliminating the torsion field, still a basic question arises: what happens to the proposed scenario (having torsion already at the level of the Lagrangian) if simpler or general torsion contributions are considered. 
\\ Among this huge spectrum of possible Lagrangian, the first case which appears as the most important to be investigated is the so-called Holst action \cite{Ashtekar1,Barbero1,Holst:1995pc,Immirzi:1996di,Rovelli:1997na}, properly considered in the presence of an Immirzi field. In fact, this is the typical term addressed in the theory of Loop 
Quantum Gravity \cite{Thiemann:2007zz,Rovelli}, of course in the 
simpler case of an Immirzi parameter, instead of a real field.
\\ Moreover, motivated by the results in \cite{Bombacigno:2018tyw}, where it has been outlined that the Immirzi field, in the considered models, always frozen out in the late Universe expansion, we are confident that the Immirzi field here enclosed in the Holst Lagrangian for a $f(R)$ extension is to be regarded as dynamical field mainly in a local sense, while its late time cosmological value is fixed. 
\\ The presence of an Holst term in place of 
a Nieh-Yan induces a significant degree of complexity in the form of the torsion, giving raise to new remarkable effects, as the emergence of a modified Newtonian potential in the static limit. Moreover, it is still possible to accomplish a full description of the gravitational wave phenomenology \cite{Maggiore}, by giving a firm and testable signature of the model. 
\\The main simplification of leaving near a 
Minkowski space-time consists of the possibility to frozen out the scalar degree of freedom coming from the $f(R)$ Lagrangian term, whose perturbation identically vanishes. Thus, we deal with the standard modes of the Einsteinian gravitational waves and a scalar wave, associated to the perturbation of the Immirzi field. In principle these two tensor and scalar modes are decoupled from each other, but their effects in the geodesic deviation equation naturally mix, given that the two deformations are simultaneously present in the displacements of a particle array. 
In particular, we arrive to define effective polarizations of the gravitational waves, understood as modification of the natural cross and plus modes of General Relativity. 
\\The presence of the perturbed Immirzi field induces an anomalous deformation of circle of particle, as effect of the incoming gravitational wave. The plus polarization acquires, in its effective manifestation, a different eccentricity for the ellipses, in 
the two orthogonal direction, in which the ring of test masses is deformed. Instead, the cross polarization, beyond possessing an expansion mode, is characterized by an angle between the two main axes slightly different from $\pi/2$. 
\\ Such phenomenological issues are qualitatively similar to those ones observed in a theory based on a Nieh-Yan term, as in \cite{Bombacigno:2018tbo}, and the two signature are distinct in their specific morphology, so that a study of the real incoming signals could discriminate between these two models and among them and other modified theories of gravity. On the present level of sensitivity also the different character of the upper limit that we could put on these scenarios could constitute a preliminar, but intriguing feature from a theoretical point of view. 
\\We also stress that, the obtained phenomenology is intrinsically different from other modified theories of gravity, because it concerns the effect of curvature on 
material device, like the actual interferometers LIGO and VIRGO \cite{TheLIGOScientific:2016src,Abbott:2017tlp,Abbott:2018utx}, more than intrinsic modification of the space-time distances. In fact, typically, in $f(R)$ theories, 
the modified polarization modes are intrinsically deformed with respect to the standard General Relativity ones. 
This distinction can have implications on the techniques of data analysis, but overall can trigger the formulation of new detector, able to distinct between these two independent modified features, \textit{i.e.} intrinsic and effective ones.
\\ The paper is structured as follows. In Sec.~\ref{sec:The Holst-$f(R)$ models} we give a detailed description of the model considered in terms of a scalar-tensor formulation, showing the equations of motion and the form of the torsion; in Sec.~\ref{sec: Linearized theory} we discuss the linear framework of the theory, outlining the freezing of the degree of freedom associated to the $f(R)$ function; in Sec.~\ref{sec:Modified Newtonian potential} we study the model in the Newtonian limit and we infer the existence of a modified gravitational potential; in Sec.~\ref{sec:Holst signature} we investigate the effects of a dynamical Immirzi field on standard gravitational polarizations, stressing the emergence of anomalous modes; finally, in Sec.~\ref{sec:concluding remarks} conclusions are drawn.

\section{The Holst-$f(R)$ models}\label{sec:The Holst-$f(R)$ models}
The starting point of our analysis is the following extension of the Holst action in vacuum\footnote{We set $c=1$}:
\begin{equation}
S=\frac{1}{16\pi G}\int d^4x\;\sqrt{-g}\leri{f(R)-\frac{\beta(x)}{2}\epsilon^{\mu\nu\rho\sigma}R_{\mu\nu\rho\sigma}},
\label{Haction}
\end{equation}
where $\beta(x^{\mu})$ is the reciprocal of the Immirzi field \cite{Ashtekar1,Barbero1,Holst:1995pc,Immirzi:1996di}, that couples to the Riemann tensor by means of the completely antisymmetric tensor (Holst term). We point out that with respect to the standard approach in LQG where it is a free parameter ruling a canonical transformation in the phase space, in our treatment it is promoted to be a real scalar field \cite{Bombacigno:2016siz}. The function $f(R)$ depends on the Ricci scalar $R$, which reads as:
\begin{equation}
R=g^{\mu\nu}R_{\mu\nu}(\Gamma)=g^{\mu\nu}R\indices{^{\rho}_{\mu\rho\nu}}(\Gamma),
\label{RicciP}
\end{equation}
where the Riemann tensor $R_{\mu\nu\rho\sigma}$ is given by:
\begin{equation}
R^{\mu}_{\;\;\nu\rho\sigma}=\partial_{\rho}\Gamma^{\mu}_{\;\;\nu\sigma}-\partial_{\sigma}\Gamma^{\mu}_{\;\;\nu\rho}+\Gamma^{\mu}_{\;\;\tau\rho}\Gamma^{\tau}_{\;\;\nu\rho}-\Gamma^{\mu}_{\;\;\tau\sigma}\Gamma^{\tau}_{\;\;\nu\rho}.
\end{equation}
We stress the fact that $\Gamma\indices{^\rho_{\mu\nu}}$ is an independent variable with respect to metric and it can be formally decomposed in:
\begin{equation}
\Gamma^{\rho}_{\;\;\mu\nu}=\bar{\Gamma}^{\rho}_{\;\;\mu\nu}+K^{\rho}_{\;\;\mu\nu},
\label{GK}
\end{equation}
where $\bar{\Gamma}^{\rho}_{\;\;\mu\nu}$ represents the well-known Levi-Civita connection\footnote{Torsionless quantities are specified by an upper bar.}, which depends on the metric variable and its derivatives only, and $K^{\rho}_{\;\;\mu\nu}$ the contorsion tensor, related to the torsion via
\begin{equation}
K^{\rho}_{\;\;\mu\nu}=\frac{1}{2}\left(T\indices{^{\rho}_{\mu\nu}}-T\indices{_{\mu}^{\rho}_{\nu}}-T\indices{_{\nu}^{\rho}_{\mu}}\right).
\label{contorsion}
\end{equation}
By analogy with \cite{Sotiriou:2008rp}, the action \eqref{Haction} can be rewritten as:
\begin{equation}
S=\frac{1}{2}\int d^4x\;\sqrt{-g}\leri{\varphi R-V(\varphi)-\frac{\beta(x)}{2}\epsilon^{\mu\nu\rho\sigma}R_{\mu\nu\rho\sigma}},
\label{Haction_phi}
\end{equation}
where $\varphi\equiv f'(R)>0$ and $V(\varphi)\equiv \varphi R(\varphi)-f(R(\varphi))$.
\\ Now, following the analysis made in \cite{Calcagni:2009xz}, it is possible to split the torsion tensor into its irreducible representations
\begin{equation}
T_{\mu\nu\rho}=\frac{1}{3}\left(T_{\nu}g_{\mu\rho}-T_{\rho}g_{\mu\nu}\right)-\frac{1}{6}\epsilon_{\mu\nu\rho\sigma}S^{\sigma}+q_{\mu\nu\rho}
\label{torsion}
\end{equation}
being $T_{\mu}=T^{\nu}_{\;\;\mu\nu}$ the trace vector, $S_{\sigma}=\epsilon_{\mu\nu\rho\sigma}T^{\mu\nu\rho}$ the pseudo-trace axial vector and $q_{\mu\nu\rho}$ the completely antisymmetric traceless component ($q^{\mu}_{\;\;\nu\mu}=0,\;\epsilon_{\mu\nu\rho\sigma}q^{\mu\nu\rho}=0$).
\\ Then, taking into account \eqref{contorsion}, if we solve the equations of motion stemming from \eqref{Haction_phi} for the torsion components \eqref{torsion}, we get (see always \cite{Calcagni:2009xz} for comparison)
\begin{equation}
\begin{split}
&T_{\mu}=\frac{3}{2(\varphi^2+\beta^2)}\leri{\varphi\overline{\nabla}_{\mu}\varphi+\beta\overline{\nabla}_{\mu}\beta}\\
&S_{\mu}=\frac{6}{\varphi^2+\beta^2}\leri{\varphi\overline{\nabla}_{\mu}\beta-\beta\overline{\nabla}_{\mu}\varphi}\\
&q_{\mu\nu\rho}=0,
\end{split}
\label{torsion_sol}
\end{equation}
where $\overline{\nabla}$ denotes the torsion-less covariant derivative with respect to Levi-Civita connection.  Eventually, by means of \eqref{torsion_sol} the contorsion tensor can be expressed in terms of $\varphi$ and the Immirzi field $\beta$, that is:
\begin{equation}
\begin{split}
K_{\mu\nu\rho}=&\frac{1}{(\varphi^2+\beta^2)}\leri{g_{\rho [\mu}\leri{\varphi\overline{\nabla}_{\nu]}\varphi+\beta\overline{\nabla}_{\nu]}\beta}}\\
&-\frac{1}{2(\varphi^2+\beta^2)}\epsilon_{\mu\nu\rho\sigma}\leri{\varphi\overline{\nabla}^{\sigma}\beta-\beta\overline{\nabla}^{\nu}\varphi},
\end{split}
\label{contorsion_sol}
\end{equation}
where square brackets denote anti-symmetrization on the indices, namely $A_{[\mu\nu]}\equiv 1/2 (A_{\mu\nu}-A_{\mu\nu})$.
\\ Now, plugging \eqref{contorsion_sol} in \eqref{GK} (see \cite{Bombacigno:2018tbo}), it is possible to recast \eqref{Haction_phi} in the more suitable form:
\begin{equation}
S_J=\frac{1}{2}\int d^4x\;\sqrt{-g}\leri{\varphi \bar{R}+g^{\mu\nu}\Xi_{\mu\nu}(\varphi,\beta)-V(\varphi)},
\label{Scalartensor}
\end{equation}
with
\begin{equation}
\begin{split}
\Xi_{\mu\nu}(\varphi,\beta)\equiv &-\frac{3\varphi}{2\leri{\varphi^2+\beta^2}}\leri{\overline{\nabla}_{\mu}\beta\overline{\nabla}_{\nu}\beta-\overline{\nabla}_{\mu}\varphi\overline{\nabla}_{\nu}\varphi}+\\
&+\frac{3\beta}{\varphi^2+\beta^2}\overline{\nabla}_{\mu}\varphi\overline{\nabla}_{\nu}\beta,
\end{split}\label{def_Xi}
\end{equation}
and $\bar{R}$ denoting the torsion-less Ricci scalar depending on Levi-Civita connection only.
Hence, varying \eqref{Scalartensor} with respect to the metric field $g_{\mu\nu}$ yields
\begin{equation}
\begin{split}
\bar{R}_{\mu\nu}-\frac{1}{2}g_{\mu\nu}\bar{R}=&-\frac{\Xi_{\mu\nu}(\varphi,\beta)}{\varphi}\;+\\&+\frac{1}{2}g_{\mu\nu}\leri{\frac{\Xi_{\;\;\rho}^{\rho}(\varphi,\beta)-V(\varphi)}{\varphi}}+\\
&+\frac{1}{\varphi}\leri{\overline{\nabla}_{\mu}\overline{\nabla}_{\nu}\varphi-g_{\mu\nu}\overline{\Box}\varphi},
\label{eq_grav}
\end{split}
\end{equation}
whereas the equations for the scalar fields $\varphi$ and $\beta$ turn out to be, respectively:
\begin{equation}
\begin{split}
\bar{R}=&+\frac{3\leri{\varphi^2-\beta^2}}{2\leri{\varphi^2+\beta^2}^2}\leri{\overline{\nabla}_{\mu}\beta\overline{\nabla}^{\mu}\beta-\overline{\nabla}_{\mu}\varphi\overline{\nabla}^{\mu}\varphi}\\
&-\frac{6\varphi\beta}{\leri{\varphi^2+\beta^2}^2}\overline{\nabla}_{\mu}\varphi\overline{\nabla}^{\mu}\beta+\frac{3}{\varphi^2+\beta^2}\leri{\varphi\overline{\Box}\varphi+\beta\overline{\Box}\beta}\\
&+V'(\varphi)
\label{eq_phi}
\end{split}
\end{equation}
where a prime represents differentiation with respect to the argument and
\begin{equation}
\begin{split}
\varphi\overline{\Box}\beta-\beta\overline{\Box}\varphi=&+\frac{\varphi\beta}{\varphi^2+\beta^2}\leri{\overline{\nabla}_{\mu}\beta\overline{\nabla}^{\mu}\beta-\overline{\nabla}_{\mu}\varphi\overline{\nabla}^{\mu}\varphi}\\
&+\frac{\varphi^2}{\varphi^2+\beta^2}\overline{\nabla}_{\mu}\varphi\overline{\nabla}^{\mu}\beta.
\label{eq_beta}
\end{split}
\end{equation}
Even though $\beta(x)$ couples directly to the Riemann tensor in \eqref{Haction}, we note that the equation for the Immirzi field obtained from the effective action \eqref{Scalartensor} does not represent a constraint for the gravitational degrees of freedom, but an highly  non-trivial relation between $\beta$ and $\varphi$. Moreover, we stress the fact that when we set $\beta(x)=0$ we are not recovering standard Palatini $f(R)$ formulation, since now torsion is allowed to be present. Indeed, for a vanishing Immirzi field, action \eqref{Scalartensor} is still equivalent to a Brans-Dicke theory of parameter $\omega=-3/2$ like the ordinary scalar-tensor representation of Palatini $f(R)$ models, but the kinetic term for $\varphi$ is now actually due to \eqref{torsion_sol} and it does not stem from the well-known Levi-Civita solution for the affine connection in terms of the conformally rescaled metric $\tilde{g}_{\mu\nu}=f'(R)g_{\mu\nu}$ (see \cite{Sotiriou:2008rp}).\\ \noindent Then, combining the trace of \eqref{eq_grav} with \eqref{eq_phi} and \eqref{eq_beta}, we can obtain the modified structural equation:
\begin{equation}
2V(\varphi)-\varphi V'(\varphi)=\frac{3\beta^3}{\leri{\varphi^2+\beta^2}^2}\overline{\nabla}_{\mu}\varphi\overline{\nabla}^{\mu}\beta,
\label{trace}
\end{equation}
and we point out that with respect to the usual Palatini vacuum case ($T_{\mu\nu}=0$) the R.H.S. of \eqref{trace} is not vanishing, but a coupling between the Immirzi field and $\varphi$ arises. Though, in principle, it allows us to solve $\varphi$ in terms of $\beta$ once we chose a specific $f(R)$ model (see \cite{Bombacigno:2018tbo,Bombacigno:2018tyw}), to accomplish such a purpose is not in general feasible. Therefore, the weak field limit represents a simpler and not less significant context where the effects of the Immirzi field could be studied, regarding both the static (Newtonian) case and the gravitational waves propagation.
\section{Linearized theory}
\label{sec: Linearized theory}
Let us consider the metric perturbation around the Minkowski background
\begin{equation}
g_{\mu\nu}=\eta_{\mu\nu}+h_{\mu\nu},
\end{equation}
being $|h_{\mu\nu}|\ll 1$ valid in some reference frame, $\eta_{\mu\nu}=diag(-1,1,1,1)$ and the inverse metric given by
\begin{equation}
g^{\mu\nu}=\eta^{\mu\nu}-h^{\mu\nu},
\end{equation}
in order to $g^{\mu\rho}g_{\rho\nu}=\delta^\mu_\nu+\mathcal{O}(h^2)$ be preserved.
At the first order in $h_{\mu\nu}$ the torsion-free Riemann tensor and the Ricci tensor read as, respectively:
\begin{align}
\bar{R}_{\rho\sigma\mu\nu}&=\frac{1}{2}\leri{\partial_{\sigma}\partial_{\mu}h_{\rho\nu}+\partial_{\rho}\partial_{\nu}h_{\sigma\mu}-\partial_{\sigma}\partial_{\nu}h_{\rho\mu}-\partial_{\rho}\partial_{\mu}h_{\sigma\nu}}\label{Riemann_lin}\\
\bar{R}_{\mu\nu}&=\frac{1}{2}\leri{\partial_{\mu}\partial_{\rho}h\indices{^{\rho}_{\nu}}+\partial_{\nu}\partial_{\rho}h\indices{^{\rho}_{\mu}}-\partial_{\mu}\partial_{\nu}h-\bar{\Box}h_{\mu\nu}}\label{Ricci_ten_lin},
\end{align}
where the trace $h$ is defined as $h\equiv \eta^{\mu\nu}h_{\mu\nu}$. Then, by virtue of \eqref{Ricci_ten_lin}, the Ricci scalar turns out to be
\begin{equation}
\bar{R}=\eta^{\mu\nu}\bar{R}_{\mu\nu}=\partial_{\mu}\partial_{\nu}h^{\mu\nu}-\bar{\Box}h.
\label{Ricci_sc_lin}
\end{equation}
Analogously, let us expand the fields $\phi$ and $\beta$ as
\begin{equation}
\phi=\phi_0+\delta\phi\quad\quad\beta=\beta_0+\delta\beta,
\end{equation}
with $\delta\beta,\,\delta\phi$ of order $\mathcal{O}(h)$, and where the background values $\beta_0$ and $\phi_0$ are determined requiring that Minkowski metric $\eta_{\mu\nu}$ be a solution for \eqref{eq_grav}-\eqref{eq_beta}. Especially, it is easy to see that they have to satisfy the following set of equations
\begin{equation}
\begin{cases}
&\frac{2V(\phi_0)}{\phi_0}\frac{\phi_0^2-\beta_0^2}{\phi_0^2+\beta_0^2}-V'(\phi_0)=0\\
&\beta_0 V(\phi_0)=0,
\end{cases}
\end{equation}
which, for the specific case $\beta_0=0$, simply constraints $\phi_0$ to be a root of the structural equation \eqref{trace}. However, since we want to keep $\beta_0$ generic as we are interested in recovering, eventually, the standard LQG limit where $\beta$ is a constant, the requirement of $V(\phi_0)=0$ leads $\phi_0$ to be a steady minimum for the potential, \textit{i.e.} $V'(\phi_0)=0$,
that into a neighbourhood of $\phi_0$ can be then approximated by \cite{Corda}
\begin{equation}
V(\phi)\simeq \frac{1}{2}m^2\delta\phi^2\Rightarrow
V'(\phi)\simeq m^2\delta\phi,
\label{Holst_f(R)_V_pert}
\end{equation}
being $m$ a positive constant.
Now, if we plug these results in \eqref{trace}, since the R.H.S. is of order $\mathcal{O}(h^2)$ in perturbation, the local fluctuation $\delta\varphi$ is compelled to vanish at the lowest order. Consequently, the gravitational equations \eqref{eq_grav} reduce to the standard vacuum case of General Relativity, \textit{i.e.}
\begin{equation}
\bar{R}_{\mu\nu}-\frac{1}{2}\eta_{\mu\nu}\bar{R}\simeq 0,
\label{eq_grav_vac}
\end{equation}
being all the terms depending on $\delta\beta$ of order $O(h^2)$, while the equation for the Immirzi field decouples from $\delta\varphi$ and simply turns out to be a vacuum wave equation for a scalar field, that is
\begin{equation}
\overline{\Box}\delta\beta\simeq 0.
\label{eq_beta_vac}
\end{equation}
Such an outcome seems to suggest that in the presence of a non-minimal coupling as in \eqref{Haction} between $\beta(x)$ and the gravitational field, the condition $\beta_0\neq 0$ leads us to a quite different extension, according the Palatini approach, of $f(R)$ theories. In fact, result \eqref{eq_grav_vac} points out that deviations form standard General Relativity predictions are of higher order with respect to the standard formulation \cite{Olmo:2005hd,Olmo:2011uz} and it is reasonable that they could appear as next-to-leading order corrections. Obviously, if we disregard the requirement $\beta_0\neq 0$, the first term of the potential expansion $V_0,\,V'_0$ do not have to vanish and the structural equation \eqref{trace} admits solutions for $\delta\varphi\neq 0$, partially restoring the well-established results in literature \cite{Olmo:2011uz}, even though we still expect non-negligible effects from the dynamics of $\delta\beta$.
\\ \indent We note that \eqref{eq_grav_vac} is in agreement with \cite{Bombacigno:2018tbo}, and also in the presence of the Holst term the gravitational field equation can be rearranged in the know form
\begin{equation}
\overline{\Box}\tilde{h}_{\mu\nu}\simeq 0,
\label{eq_grav_wave}
\end{equation}
where we introduced the trace-reverse tensor
\begin{equation}
\tilde{h}_{\mu\nu}\equiv h_{\mu\nu}-\frac{1}{2}\eta_{\mu\nu}h^{\rho}_{\;\rho},
\label{trace-reverse}
\end{equation}
and the ordinary Lorentz gauge ($\partial_{\mu}\tilde{h}^{\mu\nu}=0$) and traceless condition ($h^{\rho}_{\;\rho}=0$) have been imposed. Therefore, it follows from \eqref{eq_grav_wave} that no additional polarization is predicted for the gravitational waves and we just retain the classical plus and cross modes \cite{Rizwana:2016qdq,Berry:2011pb}. However, as already pointed out in the analogous Nieh-Yan case, in general we expect that the dynamics of the Immirzi field as described by \eqref{eq_beta_vac} could affect significantly the detection of the standard gravitational waves. \\ Indeed, in order to see that, let us evaluate the contorsion tensor \eqref{contorsion_sol} within the linearized frame, \textit{i.e.}
\begin{equation}
\begin{split}
K_{\mu\nu\rho}\simeq&\frac{\beta_0}{2(\varphi^2_0+\beta^2_0)}\leri{g_{\mu\rho}\overline{\nabla}_{\nu}\delta\beta-g_{\nu\rho}\overline{\nabla}_{\mu}\delta\beta}+\\
&-\frac{\varphi_0}{2(\varphi^2+\beta^2)}\epsilon_{\mu\nu\rho\sigma}\overline{\nabla}^{\sigma}\delta\beta.
\end{split}
\label{contorsion_pert}
\end{equation}
We stress the fact that with respect to Nieh-Yan formulation, the contorsion tensor is not completely anti-symmetric: A term proportional to $\beta_0$ arises, just anti-symmetric into the first two indices. For that reason, before dealing with the effects of the Immirzi field on the gravitational waves propagation, it can be enlightening to face the Newtonian limit of the theory, with the aim of seeking for modifications to the gravitational potential due to $\delta\beta$, as emerging from the geodesics equation analysis.

\section{Modified Newtonian potential}
\label{sec:Modified Newtonian potential}
In ordinary General Relativity, for the static weak field case the metric line element can be written as \cite{Carroll:2004st}:
\begin{equation}
ds^2=-(1+2\Psi)dt^2+(1-2\Psi)(dx^2+dy^2+dz^2),
\label{New_metric}
\end{equation}
where $\Psi$ represents the Newtonian potential, described in vacuum by the Laplace equation $\overline{\nabla}^2\Psi=0$ and related to the metric perturbation by $h_{00}=h_{ii}=-2\Psi$.
\\ We remark the fact that by considering \eqref{New_metric} we are using \eqref{eq_grav_vac}, that allows us to pick for the metric perturbations $h_{00}$ and $h_{ii}$ the same potential $\Psi$, \textit{i.e.} we are fixing the PPN parameter $\gamma=1$ \cite{Blanchet:2013haa,PoissonWill}.
\\ Then, let us consider the auto-parallel equation
\begin{equation}
\frac{d^2x^{\mu}}{d\tau^2}+\Gamma\indices{^\mu_{\rho\sigma}}\frac{dx^{\rho}}{d\tau}\frac{dx^{\sigma}}{d\tau}=0,
\label{autoparallel}
\end{equation}
with $\Gamma\indices{^\mu_{\rho\sigma}}$ now given by \eqref{GK}. Because of the symmetry properties of $K\indices{^\mu_{\rho\sigma}}$ and the independence of $\delta\beta$ on time, the only non-vanishing component of \eqref{autoparallel} if for $\mu\neq 0$, namely\footnote{In the Newtonian limit the following relations hold:
\begin{equation}
\frac{dx^i}{d\tau}\ll1,\quad\frac{dx^i}{d\tau}\ll\frac{dx^0}{d\tau}
\end{equation}}
\begin{equation}
\frac{d^2x^{i}}{d\tau^2}+\leri{\bar{\Gamma}\indices{^i_{00}}+K\indices{^i_{00}}}\leri{\frac{dx^0}{d\tau}}^2=0,
\label{autoparallel_Newt}
\end{equation}
with
\begin{equation}
\bar{\Gamma}\indices{^i_{00}}=\partial_i\Psi,\quad K\indices{^i_{00}}=\frac{\beta_0}{2\leri{\varphi_0^2+\beta_0^2}}\partial_i\delta\beta.
\end{equation}
Conveniently rescaling the time coordinate, relation \eqref{autoparallel_Newt} can be recast as
\begin{equation}
\frac{d^2x^i}{dt^2}=-\partial_i\Psi_{(\beta)},
\label{autoparalle_mod}
\end{equation}
that represents the equation for the modified Newtonian potential $\Psi_{(\beta)}$ defined as
\begin{equation}
\Psi_{(\beta)}(r)\equiv\Psi(r)+\frac{\beta_0\delta\beta(r)}{2\leri{\varphi_0^2+\beta_0^2}},
\end{equation}
where the radial coordinate is defined as $r\equiv\sqrt{x_i x^i}$.
\\ Given that \eqref{eq_beta_vac} in the static limit reduces to $\overline{\nabla}^2\delta\beta\simeq 0$, a solution for \eqref{autoparalle_mod} can be easily found, \textit{i.e.}:
\begin{equation}
\Psi_{(\beta)}(r)=-\frac{GM}{r}+\frac{\beta_0 A}{2(\varphi_0^2+\beta_0^2)}\frac{1}{r}=-\frac{G_{(\beta)}M}{r}
\label{pot_mod}
\end{equation}
with $M$ the gravitational mass of the source and $G_{(\beta)}$ an effective Newtonian constant, defined as $G_{(\beta)}\equiv G(1-\epsilon)$. In particular, we introduced the parameter $\epsilon\equiv\frac{\beta_0 C}{2(\varphi_0^2+\beta_0^2)}$, with the integration constant for $\delta\beta$ rewritten as $A=C\cdot GM$ with C a dimension-less factor\footnote{We imposed that $\delta\beta$ be asymptotically vanishing.}.
\\ Now, assuming that at the lowest order the orbits around the Sun could be considered circular, by virtue of \eqref{pot_mod} it is possible to evaluate the modified orbital period, that is
\begin{equation}
T_{(\beta)}=2\pi\sqrt{\frac{r^3}{G_{(\beta)}M}}.
\label{period_mod}
\end{equation}
The parameter $\epsilon$, that takes account for the deviation from classical predictions,  can be constrained comparing \eqref{period_mod} with the Keplerian expression $T_K=2\pi r^{3/2}(GM)^{-1/2}$ and requiring that the correction be smaller that the experimental uncertainty \cite{Zakharov:2006uq,Chiba:2006jp,Schmidt:2008qi,Nojiri:2007as,Turyshev:2008dr}, \textit{i.e.}
\begin{equation}
\frac{\abs{T_K-T_{(\beta)}}}{T_K}\le\frac{\delta T_{exp}}{T_{exp}},
\end{equation}
which considering \eqref{period_mod}, yields up the first order in $\delta$ to
\begin{equation}
\abs{\epsilon}\le 2\frac{\delta T_{exp}}{T_{exp}}.
\label{delta estimate mod Newton}
\end{equation}
We conclude this section noting that if we consider the pertubation $\delta\beta$ in a local sense and we assume the value $\beta_0$ to be fixed by LQG estimates \cite{Ghosh:2004wq} (mainly from black hole entropy, even though further investigations ruled out this possibility \cite{Ghosh:2011fc}), in the limiting case of $\phi_0=1$ ($f(R)=R$) relation \eqref{delta estimate mod Newton} allows to fix directly the value of $C$, recovering a thorough description of the dynamics.

\section{Holst signature for gravitational waves}\label{sec:Holst signature}
In order to investigate the consequences of a dynamical Immirzi field on the gravitational waves propagation, let us take the geodesic deviation equation, evaluated in the comoving frame, \textit{i.e.}:
\begin{equation}
\frac{\partial^2}{\partial \tau^2}\xi^{\alpha}=R\indices{^{\alpha}_{\mu\nu\beta}}u^{\mu}u^{\nu}\xi^{\beta}=R\indices{^{\alpha}_{00\beta}}\xi^{\beta},
\label{geodesic}
\end{equation}
with $\xi^
\alpha=(0,\xi_x,\xi_y,\xi_z)$ a vector denoting the separation between two nearby geodesics and the Riemann tensor given, up to the first order, by:
\begin{equation}
R\indices{^{\rho}_{\mu\sigma\nu}}=\bar{R}\indices{^{\rho}_{\mu\sigma\nu}}+\partial_{\sigma}K\indices{^{\rho}_{\mu\nu}}-\partial_{\nu}K\indices{^{\rho}_{\mu\sigma}}.
\label{RiemK}
\end{equation}
Then, if we consider a gravitational plane wave in the TT-gauge which propagates along the $z$ direction, namely:
\begin{equation}
\tilde{h}_{ij}(t,z)\equiv
\begin{pmatrix}
h_{+}(t,z) &  h_{\times}(t,z) \\
h_{\times}(t,z) & - h_{+}(t,z)
\end{pmatrix},
\end{equation}
where $i,j=x,y$, the only non-vanishing components of torsion-less Riemann tensor $\bar{R}\indices{^{\rho}_{\mu\sigma\nu}}$ are:
\begin{equation}
\begin{split}
&\bar{R}\indices{^x_{0x0}}=\bar{R}_{x0x0}=-\bar{R}_{y0y0}=-\frac{1}{2}\frac{\partial^2h_{+}}{\partial t^2}\\
&\bar{R}\indices{^y_{0x0}}=\bar{R}_{y0x0}=\bar{R}_{x0y0}=-\frac{1}{2}\frac{\partial^2h_{\times}}{\partial t^2}.
\end{split}
\label{hTT}
\end{equation}
Now, by virtue of \eqref{contorsion_pert} and \eqref{hTT}, equation \eqref{geodesic} yields to, for a generic $\delta\beta=\delta\beta(t,\vec{x})$ wave:
\begin{equation}
\begin{split}
&\frac{\partial^2\xi_x}{\partial t^2}=\frac{1}{2}\leri{\xi_x\frac{\partial^2h_{+}}{\partial t^2}+\xi_y\frac{\partial^2h_{\times}}{\partial t^2}}+\frac{\mathcal{D}_x(\vec{\xi};\varphi_0,\beta_0)}{2(\varphi_0^2+\beta_0^2)}\;\delta\beta\\
&\frac{\partial^2\xi_y}{\partial t^2}=\frac{1}{2}\leri{\xi_x\frac{\partial^2h_{\times}}{\partial t^2}-\xi_y\frac{\partial^2h_{+}}{\partial t^2}}+\frac{\mathcal{D}_y(\vec{\xi};\varphi_0,\beta_0)}{2(\varphi_0^2+\beta_0^2)}\;\delta\beta\\
&\frac{\partial^2\xi_z}{\partial t^2}=\frac{\mathcal{D}_z(\vec{\xi};\varphi_0,\beta_0)}{2(\varphi_0^2+\beta_0^2)}\;\delta\beta,
\end{split}
\label{geodesicimmirzi}
\end{equation}
where $\mathcal{D}_k(\vec{\xi};\varphi_0,\beta_0)$ is a second order differential operator defined by:
\begin{equation}
\mathcal{D}_k(\vec{\xi};\varphi_0,\beta_0)\equiv\varphi_0\leri{\vec{\xi}\times\vec{\nabla}}_{k}\frac{\partial}{\partial t}-\beta_0\leri{\vec{\xi}\cdot\vec{\nabla}}\frac{\partial}{\partial k}+\beta_0\xi_{k}\frac{\partial^2}{\partial t^2},
\end{equation}
where the index $k$ runs over $x,y,z$.
\\ We note that when $\delta\beta$ is aligned to the $k$-direction, the action of the corresponding operator $\mathcal{D}_k$ reduces to
\begin{equation}
\mathcal{D}_k(\vec{\xi};\varphi_0,\beta_0)\delta\beta(t,k)=-\beta_0\xi_k\overline{\Box}\delta\beta,
\end{equation}
which vanishes identically by virtue of \eqref{eq_beta_vac}: Therefore, choosing $\delta\beta$ propagating along the $z$-direction, we can restrict our analysis to the $(x,y)$ plane.
\\ \indent Now, be $\delta\beta(t,z)$ given by $(c=1)$:
\begin{equation}
\delta\beta(t,z)=\overline{\delta\beta}\sin(\omega(t-z))
\label{betaz}
\end{equation}
and let us fix $\vec{\xi}$ as
\begin{equation}
\vec{\xi}=\leri{\xi_x^{(0)}+\delta x,\xi_y^{(0)}+\delta y,0},
\label{xipert}
\end{equation}
being $\xi_x^{(0)},\,\xi_y^{(0)}$ the initial positions and $\delta x,\,\delta y$ the displacements of order $O(h)$ induced by $\delta\beta(t,z)$. When we turn off the gravitational modes $h_+,h_{\times}$, the system \eqref{geodesicimmirzi} assumes the form:
\begin{equation}
\begin{split}
&\frac{\partial^2\delta x}{\partial t^2}\simeq-\frac{\omega^2}{2(\varphi_0^2+\beta_0^2)}\leri{\beta_0\xi_x^{(0)}-\varphi_0\xi_y^{(0)}}\delta\beta(t,z)
\\
&\frac{\partial^2\delta y}{\partial t^2}\simeq-\frac{\omega^2}{2(\varphi_0^2+\beta_0^2)}\leri{\beta_0\xi_y^{(0)}+\varphi_0\xi_x^{(0)}}\delta\beta(t,z),
\end{split}
\label{onlybeta}
\end{equation}
where according to \eqref{xipert} we neglected terms of order $O(h^2)$. Thus, if we set the time origin such that $\delta\beta=0$ at $t=0$, a solution for \eqref{onlybeta} is given by
\begin{equation}
\begin{split}
\delta x(t)&\simeq\frac{1}{2}\leri{\xi_x^{(0)}\beta_B-\xi_y^{(0)}\beta_R}\sin\omega t\\
\delta y(t)&\simeq\frac{1}{2}\leri{\xi_y^{(0)}\beta_B+\xi_x^{(0)}\beta_R}\sin\omega t,
\end{split}
\label{onlybetasol}
\end{equation}
with
\begin{equation}
\beta_B\equiv\frac{\beta_0\overline{\delta\beta}}{\varphi_0^2+\beta_0^2}\quad\quad\beta_R\equiv\frac{\varphi_0\overline{\delta\beta}}{\varphi_0^2+\beta_0^2}.
\label{breath_tors_amplit}
\end{equation}

\begin{figure}
\begin{center}
\includegraphics[width=1\columnwidth]{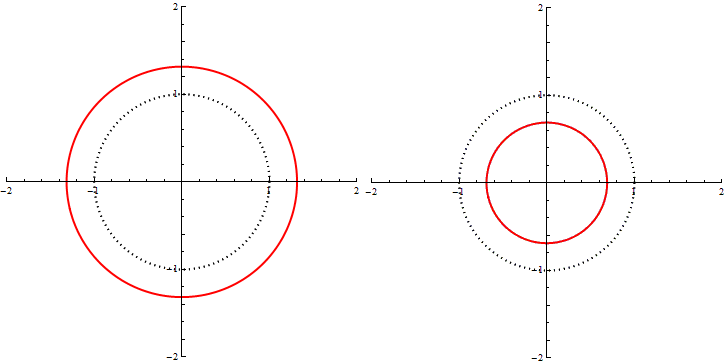}
\caption{The deformation induced by the polarization $\beta_B$ for $\omega t=\pi/2$ (left) and $\omega t=3\pi/2$ (right). The dotted circle represents the unperturbed test masses, whereas with the solid lines is shown the breathing mode. For the sake of clarity the effects due to $\delta\beta$ are magnified with respect to the actual dynamics.}
\label{BetaB_Tot}
\end{center}
\end{figure}

\begin{figure}
\begin{center}
\includegraphics[width=1\columnwidth]{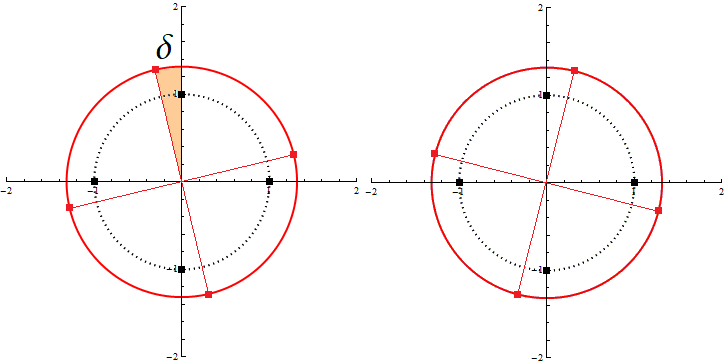}
\caption{The deformation induced by the polarization $\beta_R$ for $\omega t=\pi/2$ (left) and $\omega t=3\pi/2$ (right). The dotted circle represents the unperturbed test masses, whereas with the solid lines the $\beta_R$ mode is shown (in $\omega t=\pi,\,2\pi$ we just recover the initial rest position); the rotation angle $\delta$ is also shown. For the sake of clarity the effects due to $\delta\beta$ are magnified with respect to the actual dynamics.}
\label{BetaR_Tot}
\end{center}
\end{figure}

Whereas $\beta_B$ is responsible for a breathing mode (Fig.~\ref{BetaB_Tot}), the parameter $\beta_R$ rules a peculiar perturbation characterized by both rotation and dilation effects. Specifically, every cycle the ring of test masses is expanded and turned twice (Fig.~\ref{BetaR_Tot}) and it is worth noting that compared to a pure breathing mode, the perturbation $\beta_R$ is not endowed with a contraction phase: The test masses ring never shrinks with respect to its rest position. Eventually, a relation between the maximum rotation angle $\delta$ and the parameter $\beta_R$ can be settled, namely:
\begin{equation}
\delta=\tan^{-1}\leri{\frac{\beta_R}{2}}\simeq \frac{\beta_R}{2}.
\label{angolo}
\end{equation}

\begin{figure}
\begin{center}
\includegraphics[width=1\columnwidth]{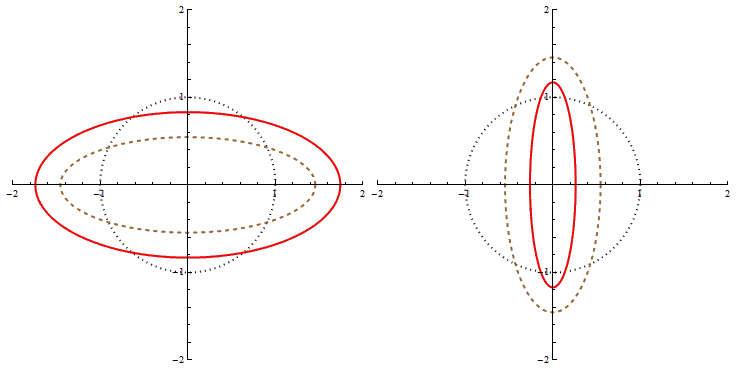}
\caption{Effective plus polarization for $\epsilon_B\ll 1$, in $\omega t=\pi/2$ (left) and $\omega t=3\pi/2$ (right). The dotted circle represents the unperturbed test masses, whereas the dashed and solid lines are the standard and modified polarization, respectively. For the sake of clarity the effects due to $\delta\beta$ are magnified with respect to the actual dynamics.}
\label{Plusmode_Tot}
\end{center}
\end{figure}

 Now, let us switch on again the gravitational modes, choosing analogously to  \eqref{betaz} $h_{+}=\overline{h_{+}}\sin\omega t$ (the same for $h_{\times}$). A solution for \eqref{geodesicimmirzi} can be formulated in terms of a pair of gravitational effective polarizations, \textit{i.e.}
\begin{equation}
\begin{split}
\delta x(t)&\simeq\frac{1}{2}\leri{\xi_x^{(0)}h^{(+)}_{+}+\xi_y^{(0)}h^{(-)}_{\times}}\sin\omega t\\
\delta y(t)&\simeq\frac{1}{2}\leri{\xi_x^{(0)}h^{(+)}_{\times}-\xi_y^{(0)}h^{(-)}_{+}}\sin\omega t,
\end{split}
\label{geodesiceff}
\end{equation}
where we introduced the modified amplitudes
\begin{equation}
h^{(\pm)}_{+}\equiv h_{+}(1\pm\epsilon_B)\quad\quad h^{(\pm)}_{\times}\equiv h_{\times}(1\pm\epsilon_R),
\label{mod_cross_plus}
\end{equation}
with $\epsilon_B\equiv\beta_B/h_{+}$ and analogously for $\epsilon_R$.
\\ In the following, without loss of generality we shall focus the analysis to $\epsilon_B,\,\epsilon_R\in (0,1)$, this choice being motivated by the request that the modifications induced by the Immirzi field on the standard polarizations be small, as outlined by the absence of observational evidences regarding anomalous modes (see \cite{TheLIGOScientific:2016src,Abbott:2017tlp,Abbott:2018utx}). Furthermore, it is always possible to extend the obtained results to the negative domain via a counter-clockwise rotation of $\pi/2$ in the $(x,y)$ plane. 
\\ Then, if we consider the effective plus mode $h^{(\pm)}_{+}$,  it is easy to see that its effect is to induce an asymmetric plus mode, characterized by stress of different strength on each axis (Fig.~\ref{Plusmode_Tot}). In particular, dilatations and contractions along the $x$ direction turn out to be larger than those ones on the $y$ axis, and the ring of test masses is distorted into ellipses of different eccentricity, given by
\begin{equation}
\begin{split}
e^{(-)}&=\sqrt{1-\leri{\frac{1-\frac{h^{(-)}_{+}}{2}\sin\omega t}{1+\frac{h^{(+)}_{+}}{2}\sin\omega t}}^2}\quad\text{for}\quad\omega t\in [0,\pi]\\
e^{(+)}&=\sqrt{1-\leri{\frac{1+\frac{h^{(+)}_{+}}{2}\sin\omega t}{1-\frac{h^{(-)}_{+}}{2}\sin\omega t}}^2}\quad\text{for}\quad\omega t\in [\pi,2\pi],
\end{split}
\label{eccentricity}
\end{equation}
where the following inqualities hold
\begin{equation}
\begin{split}
e^{(-)}&<e_{\epsilon_B=0}\quad\text{for}\quad\omega t\in [0,\pi]\\
e^{(+)}&>e_{\epsilon_B=0}\quad\text{for}\quad\omega t\in [\pi,2\pi].
\end{split}
\end{equation}

\begin{figure}
\begin{center}
\includegraphics[width=1\columnwidth]{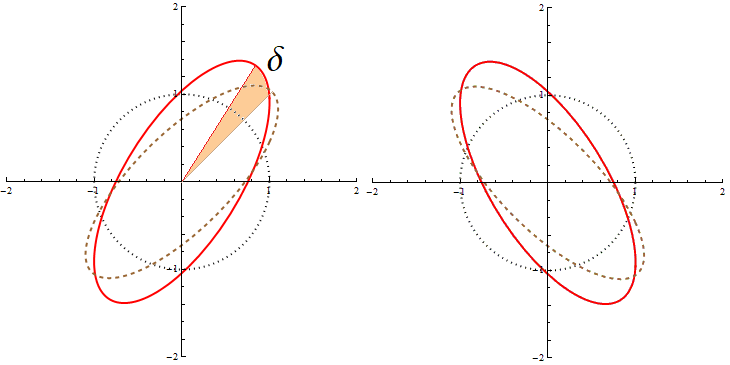}
\caption{Effective cross polarization for $\epsilon_R\ll 1$, in $\omega t=\pi/2$ (left) and $\omega t=3\pi/2$ (right). The dotted circle represents the unperturbed test masses, whereas the dashed and solid lines are the standard and modified polarization, respectively. For the sake of clarity the effects due to $\delta\beta$ are magnified with respect to the actual dynamics.}
\label{Crossmode_Tot}
\end{center}
\end{figure}
Instead, concerning the deformation due to the modified cross polarization, we note that by close analogy with what discussed previously with regard to $\beta_R$ effects, the ellipses are both rotated and enlarged with respect to standard polarization (Fig.~\ref{Crossmode_Tot}). Especially, it is worth stressing that the elongation axes turn out to be not orthogonal, but separated by the angle :
\begin{equation}
\theta=\frac{\pi}{2}-2\delta,
\end{equation}
with $\delta$ estimated by \eqref{angolo}.
\\ \indent We remark the fact that the effects predicted by \eqref{geodesicimmirzi} would be present even if we considered the particular case of $f(R)=R$. Indeed, by the inspection of \eqref{contorsion_sol} it is clear that also for $\varphi=1$ a not completely anti-symmetric component for the contorsion tensor survives, causing the appearance of both the aforementioned effective polarizations and the modified Newtonian potential, due the dynamical nature of the Immirzi field. Conversely, only if $\beta(x)$ is relaxed to a constant value the standard gravitational modes are restored, as it can be inferred by \eqref{contorsion_pert}, being always $\delta\varphi=0$ within the linearized theory.

\section{Concluding remarks}\label{sec:concluding remarks}
We analyzed a $f(R)$ extended theory of gravity in the Palatini formulation by including in the dynamics 
an Holst term, characterized by an Immirzi field. This study follows the analysis in \cite{Bombacigno:2018tbo}, 
where a similar scenario was investigated in the presence of a Nieh-Yan term, and with respect to that has a greater degree of complexity, especially in view of the possibility to eliminate the torsion field in terms of the remaining
degrees of freedom. 
\\In particular, we stressed the emergence of a modified Newtonian potential and we were able to give an estimate of the deviation from GR by comparing the modified orbital period with the Keplerian prediction.
\\Furthermore, it was possible to analyze the propagation of gravitational waves in the proposed dynamical framework, by virtue of the freezing which takes place for the scalar degree of freedom of the $f(R)$ Lagrangian, whose linear perturbation 
identically vanishes. Therefore, the propagation of the gravitational waves is the same as in General Relativity (in agreement with standard Palatini $f(R)$ formulation), but in the presence of the Immirzi field, which provides an additional scalar wave. 
However, the tensor and scalar waves simultaneously act on test particles and their combined effect can be restated as effective plus and cross standard gravitational waves. Thus, the phenomenological signature of the proposed theory is the emergence of a plus polarization anisotropically acting along the two orthogonal directions and a cross polarization which is 
characterized by a slightly modified angle with respect to $\psi /2$. Furthermore, in both cases an expansion 
effective mode is present, which enlarges and contracts the radius of a circle of particles. 
\\The basic idea, underlying this analysis like that one in \cite{Bombacigno:2018tbo}, consists of implementing a consistent Palatini formulation of the $f(R)$, by including torsion \textit{ab initio} and trying to eliminate it in term of the other dynamical field, among which an Immirzi field stands. 
\\ We have clearly demonstrated that the main signature of such restated approaches of the Palatini $f(R)$ model is 
the emergence of two effective polarization, slightly modified with respect to the two basic ones of linear General Relativity. 
A suitably setting of the data analysis for the LIGO-VIRGO incoming detections would allow to put precise upper limits, 
if not yet real measures, of the parameters governing the deformation and therefore the viability of this extended gravitational theory could be preliminary tested. In this respect, it would be relevant to set up suitable algorithms of data analysis, able to distinguish between real and effective modified gravitational wave polarizations. 
\acknowledgements{We would like to thank Fabio Moretti for the enlightening discussion about the effects of the Immirzi field on the gravitational waves polarizations.}

\end{document}